\documentclass[10pt,aps,prl,twocolumn,showpacs,superscriptaddress]{revtex4-1}  

\usepackage[modulo,switch]{lineno}
\modulolinenumbers[1]

\usepackage[utf8]{inputenc}
\usepackage{graphicx}
\usepackage{amsmath}
\usepackage{amssymb}
\usepackage{bm}        
\usepackage{txfonts}
\usepackage{multirow}
\usepackage{mathptmx}  
\usepackage{siunitx}
\usepackage{xargs}                      

\usepackage[pdftex,dvipsnames]{xcolor}  
\usepackage[normalem]{ulem} 

\usepackage[%
  colorlinks=true,
  urlcolor=blue,
  linkcolor=blue,
  citecolor=blue
]{hyperref}

\begin{document}

\title{Position measurement of a dipolar scatterer via self-homodyne detection}

\author{G.~Cerchiari}
\email{giovanni.cerchiari@uibk.ac.at}
\altaffiliation{Corresponding author}
\affiliation{Institut f\"ur Experimentalphysik, Universit\"at Innsbruck, Technikerstrasse~25, 6020~Innsbruck, Austria}
\author{L.~Dania}
\affiliation{Institut f\"ur Experimentalphysik, Universit\"at Innsbruck, Technikerstrasse~25, 6020~Innsbruck, Austria}
\author{D. S.~Bykov}
\affiliation{Institut f\"ur Experimentalphysik, Universit\"at Innsbruck, Technikerstrasse~25, 6020~Innsbruck, Austria}
\author{R.~Blatt}
\affiliation{Institut f\"ur Experimentalphysik, Universit\"at Innsbruck, Technikerstrasse~25, 6020~Innsbruck, Austria}
\affiliation{Institut f\"ur Quantenoptik und Quanteninformation, \"Osterreichische Akademie der Wissenschaften, Technikerstrasse 21a, 6020 Innsbruck, Austria}
\author{T.~Northup}
\affiliation{Institut f\"ur Experimentalphysik, Universit\"at Innsbruck, Technikerstrasse~25, 6020~Innsbruck, Austria}

\date{\today}

\begin{abstract}

We describe a technique to measure the position of a dipolar scatterer based on self-homodyne detection of the scattered light. The method can theoretically reach the Heisenberg limit, at which information gained about the position is constrained only by the back-action of the scattered light. The technique has applications in the fields of levitated optomechanics and trapped ions and is generally applicable to the position determination of confined light scatterers.
\end{abstract}
\maketitle

\section{Introduction}

Quantum mechanics is our best predictive theory to describe the microscopic world. According to this model, elementary particles are described by wave excitations of fields \cite{Peskin1995ev}, and measurements of physical quantities on these systems have probabilistic outcomes. This wave theory strongly contrasts with our everyday experience, in which macroscopic objects can be described as entities with deterministic measurable properties using classical theory. At the interface between quantum and classical scales, physicists challenge the two theories to search for new physics, for example, to understand the role of gravity in the quantum-mechanical description \cite{Aspelmeyer2014, Carlesso_2019}. 

One promising approach in these investigations is to search for quantum-mechanical properties of the motion of micron- or submicron-sized dielectric particles levitated in optical, magnetic or electrodynamic traps~\cite{aspelmeyer2012quantum, Millen_2020}. The levitation ensures a high degree of isolation in space and enables the study of the particles' wave properties stemming from the quantization of the confining potential~\cite{Felix_asimmetry,Delic2020,magrini2020}. In these systems, laser light is typically used to measure and interact with the levitated particle's motion. The standard optical approach to observe single quanta of oscillatory motion (phonons) is to detect a monochromatic light field scattered by the particle~\cite{Bullier2019, Ranjit2015, Gieseler2012}. The scattered field carries information about the location of emission but, at the same time, perturbs the momentum of the particle via photon recoil~\cite{Jain2016}. The trade-off between the information gained and the back action sets the ultimate limit for detection: the Heisenberg limit. Here, by Heisenberg limit, we mean that the product of the measurement imprecision in the determination of the scatterer position and the measurement back action on the scatterer's momentum satisfies the lower bound of the Heisenberg uncertainty relation~\cite{Clerk}. A further challenge is imposed by the angular distribution of the scattered light, which, in the Rayleigh approximation, is a dipole pattern that spreads the information over the full solid angle~\cite{hulst1981light}. 

The observation of the scattered light over the entire solid angle is the key to reaching the Heisenberg limit as proposed in Ref.~\cite{Tebbenjohanns2019}. In the Gedankenexperiment described therein, the following geometrical configuration is proposed. The scattered field is interfered with a second dipolar field having the same wavefronts and polarization, allowing for the reconstruction of the scatterer's position via homodyne detection. The secondary dipole field (reference field) must be permanently located at a reference position which defines the origin of the scatterer's displacements.

This theoretical configuration poses the challenge of generating the dipolar reference field, while simultaneously detecting light over the entire solid angle.
This optical problem has been approached in the past by locating the scatterers in the focus of deep parabolic mirrors to collimate the dipolar field into a plane wave having a Gaussian $\textrm{TEM}_{01*}$ profile, a so-called doughnut mode ~\cite{Salakhutdinov16, Salakhutdinov2020}. This solution requires the parabolic mirror to cover a large fraction of the solid angle around the scatterer to collimate the field and it may be difficult to implement in experiments with limited optical access. In this work, we propose that self-homodyne of the field emitted by the particle is a viable implementation of the ideal configuration. In a symmetric realization of the self-homodyne technique, the solid angle is split into two parts. One part is occupied by a hemispherical mirror, the other by a hemispherical pixel detector. The mirror generates a secondary image of the dipole, which interferes at the detector with the primary scattered light. Similar configurations have been explored in the past to obtain nanometric resolution in microscopy applications~\cite{Swan2003, Davis2007} and in the context of half-cavity experiments to investigate the spontaneous emission of a trapped atomic ion  \cite{Bushev2006, Bushev2013, slodivcka2012interferometric, Cerchiari2020}. The previous studies did not address the theoretical limitations for motion detection or the possibility to extend the technique beyond the field of trapped ions. Here, we prove that the self-homodyne method is equivalent to the ideal configuration, and we present how it can be used to obtain lower imprecision for position detection than obtained in state-of-the-art experiments in the field of levitated optomechanics.

\section{Self-homodyne detection and the Heisenberg limit}

\begin{figure}
\includegraphics[width=\columnwidth]{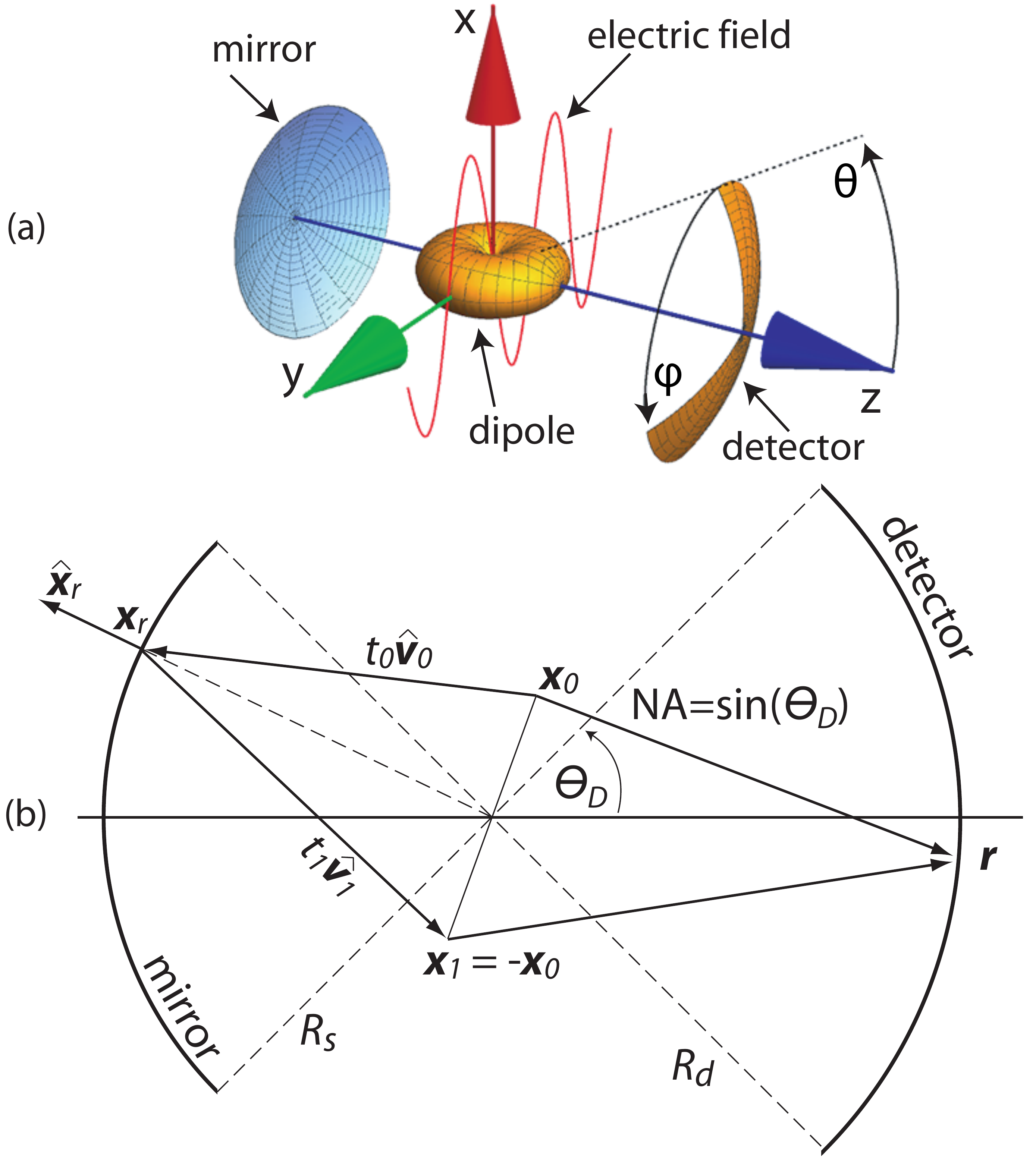}
\caption{Schematic representation of the self-homodyne setup. a) Three-dimensional overview represented in both the Cartesian coordinate system and the spherical coordinate system used in this article. b) Bi-dimensional section. A spherical mirror of radius $R_s$ images the light emitted by a dipolar scatterer located at $\bm{x}_0$ at the point $\bm{x}_1=-\bm{x}_0$. An optical path for the spherical wavefronts is drawn departing from $\bm{x}_0$, reflected at $\bm{x}_r$, and imaged at $\bm{x}_1$. The ray travels an optical path given by $t_0+t_1=2R_s$ along the directions defined by the unitary vectors (versors) $\hat{\bm{v}}_0$ and $\hat{\bm{v}}_1$. The light emitted by the scatterer interferes with the scatterer's image on the detector of radius $R_d$ at position $\bm{r}$. The detector collects all the light within the polar angle interval $\left[0, \theta_D\right]$ from the optical axis bisecting the mirror and the detector. The numerical aperture of the setup is defined as $\textrm{NA}=\sin{\theta_D}$.}
\label{fig:halfcavity_drawing}
\end{figure}

A schematic drawing of the self-homodyne setup is presented in Fig.~\ref{fig:halfcavity_drawing}. The mirror of radius $R_s$ and the detector of radius $R_d$ are concentric around the origin. A point-like dipolar emitter of radiation with wavelength $\lambda$ is located at $\bm{x}_0$ near the origin. This configuration is referred to as a ``half-cavity''~\cite{Dorner2002}. We are interested in deducing the position $\bm{x}_0$ of the scatterer with respect to the origin of the coordinate system by measuring the light intensity impinging on the detector at different angles. For the discussion, we will refer to $\bm{x}_0$ both as the position or the displacement to be detected.

We assume that the scattered light is reflected and detected in the far field. This approximation corresponds to the condition under which the modulus of the displacement $\lVert\bm{x}_0\rVert$ from the origin is much smaller than the distance $R_d$ between the particle and the detector and much smaller than the radius of curvature $R_s$ of the mirror. The far-field approximation allows us to consider the dipolar scatterer as a source of spherical waves with polarization orthogonal to the radial versor $\hat{\bm{n}}$, which indicates a generic direction in space departing from the origin. Under these conditions, we can model the light field's propagation and interference by using the Huygens–Fresnel integral in scalar form. Furthermore, the spherical optical elements enable us to solve the problem without using the paraxial approximation while disregarding the $1/R$ factor in the Green's function of the Huygens–Fresnel integral. 

At the location $\bm{r}=\hat{\bm{n}}R_d$ on the detector, the expression for the field's amplitude is:
\begin{equation}\label{eq:direct_electric}
    E_s\left(\bm{r}\right)=E_0\exp{\left(i\frac{2\pi}{\lambda}\hat{\bm{n}} \cdot \bm{x}_0 \right)} \; ,
\end{equation}
where $E_0$ is a constant complex amplitude and ``$\cdot$'' denotes the scalar product. The amplitude $E_0$ contains the re-scaling and phase pre-factor $\exp\left(-i 2 \pi R_d/\lambda\right)/R_d$, which is constant across the detector surface. For small displacement in any direction ($\lVert\bm{x}_0\rVert \ll R_s$), the mirror generates an image of the emitter at the opposite side of the origin, at $\bm{x}_1 = -\bm{x}_0$. The polarization and wavefronts of the image are the same as for the primary dipole. The field of the image has a constant phase shift corresponding to an optical path of twice the radius $R_s$ of the mirror (see Appendix A). The image field is phase-referenced to the scattered field and can be expressed at the detector as:
\begin{equation}\label{eq:image_electric}
    E_i\left(\bm{r}\right)=-\rho E_0 \exp{\left(-i\frac{2\pi}{\lambda} \left(\hat{\bm{n}} \cdot \bm{x}_0 + 2 R_s\right)\right)} \; ,
\end{equation}
where $\rho$ is the reflection coefficient for the electric field at the mirror. The minus sign in front accounts for the phase difference accumulated at the reflection.

The intensity measured at $\bm{r}$ is:
\begin{equation}\label{eq:interference_halfcavity}
    \begin{split}
    &I\left(\bm{r}\right) \propto \lvert E_s\left(\bm{r}\right) + E_i\left(\bm{r}\right)\rvert^2\\
    &\propto 1 + \rho^2 - 2\rho \cos{\left(\frac{4 \pi}{\lambda} \left( \hat{\bm{n}} \cdot \bm{x}_0+R_s\right)\right)} \; .
    \end{split}
\end{equation}
We see that the detected light intensity at any position $\bm{r}$ is sinusoidal as a function of the position $\bm{x}_0$. To detect the position via an intensity modulation univocally, the modulus of the displacement must be sub-wavelength: $\lVert \bm{x}_0\rVert<\lambda$. The expression $\lVert \bm{x}_0\rVert\ll\lambda$ is normally valid in levitated optomechanics experiments aiming to observe the oscillations of a scatterer close to the motional ground state \cite{Millen_2020,Chang1005}.

We obtain the maximum sensitivity on the slope of the interference pattern, i.e., when the equality $4\pi R_s/\lambda = 2\pi m\pm\pi/2$ holds, with $m$ a natural number. By simplifying Eq.~\ref{eq:interference_halfcavity} to first order in $\lVert \bm{x}_0\rVert/\lambda$ under these conditions, we obtain:
\begin{equation}\label{eq:int_taylor_prop}
    I\left(\bm{r}\right) \propto 1+\rho^2 \pm \rho \frac{8 \pi}{\lambda} \left(\hat{\bm{n}} \cdot \bm{x}_0\right) \; .
\end{equation}
In this formula, either sign in the last term leads to the same sensitivity. For simplicity, we select the positive sign.

To impose a stable phase of interference for maximal motion sensitivity, the experimenter can take advantage of the mass difference between the scatterer and the mirror. As already demonstrated in Refs.~\cite{Bushev2006, Cerchiari2020}, the oscillations of the particle are usually much faster than the drifts of the mirror position. By averaging the detected light over several periods of the particle's oscillation, one can generate a feedback signal to compensate for the mirror movements without compromising the information obtained about the particle's motion.

Equation~\ref{eq:int_taylor_prop} can be compared to the intensity measured for the ideal homodyne scheme, adapted from Ref.~\cite{Tebbenjohanns2019}:
\begin{equation}\label{eq:interference_ideal}
    I_{\textrm{ideal}}\left(\bm{r}\right) \propto 1 + \gamma^2 + \gamma \frac{4 \pi}{\lambda} \left(\hat{\bm{n}} \cdot \bm{x}_0\right) \; .
\end{equation}
Here, $\gamma$ is the proportionality coefficient between the reference electric field and the scattered electric field. We see that $\gamma$ substitutes for the reflection coefficient $\rho$.

The experimenter's control over the power of the reference field is manifested in the coefficient $\gamma$. In state-of-the-art experiments adopting homodyne detection schemes~\cite{tebbenjohanns2021quantum,magrini2020}, this control allows one to work in a regime in which $\gamma^2+1\sim\gamma^2$, corresponding to adjusting the intensity of the reference field to be much stronger than that of the direct field. However, this regime is not accessible for the self-homodyne method because $\rho$ is bounded by the condition $\rho \leq 1$. Furthermore, we note that the position-dependent terms in Eqs.~\ref{eq:int_taylor_prop} and \ref{eq:interference_ideal} differ by a factor of two. The discrepancy stems from the fact that, in contrast to the ideal detection configuration, the reference field source, which is the image created by the mirror, is not fixed to the center of the curvature of the mirror. The displacement of the particle causes a shift of
the image in the opposite direction relative to the center of the curvature, thus increasing the sensitivity by a factor of two. Also in contrast to the ideal case, in the self-homodyne method the half solid angle occupied by the mirror cannot be used for detection and the interference is observed only on the opposite half of the solid angle. In the remainder of this section, we calculate the limit for position detection to show that the self-homodyne setup can be considered equivalent to the ideal configuration notwithstanding these differences. In Appendix B, we expand our discussion of the self-homodyne process to highlight the directionality of the interference effect and its connection to the spontaneous emission of an atomic scatterer-mirror system for a deeper understanding of the physical phenomenon.

A dipolar scatterer is not a perfect source of spherical waves. In the far-field approximation, we can correct this model by retaining spherical wavefronts and by modifying the angular distribution of the radiated light power. The differential power $dp_{\textrm{dip}}$ per unit solid angle $d\Omega$ radiated by a dipole depends on the polarization $\hat{\bm{\epsilon}}$ of the driving light beam according to the far-field pattern~\cite{hulst1981light}:
\begin{equation}\label{eq:power_dipole}
     dp_{\textrm{dip}}=\frac{3}{8\pi}P\left(1-\lvert\hat{\bm{n}}\cdot\hat{\bm{\epsilon}}\rvert^2\right)d\Omega \; ,
\end{equation}
where $P$ is the total power radiated by an electric field with amplitude $E_0$ (see Eqs.~\ref{eq:direct_electric} and \ref{eq:image_electric}). Since the mirror is in the far field, the polarization of the image is the same as the polarization of the direct dipole. Combining the interference term and the angular distribution of the scattered radiation, we obtain the differential detected power $dp_{\textrm{det}}$:
\begin{equation}\label{eq:dpdet}
     dp_{\textrm{det}}=\left(1 + \rho^2 + \rho \frac{8 \pi}{\lambda} \left(\hat{\bm{n}}\cdot\bm{x}_0\right)\right)dp_{\textrm{dip}} \; .
\end{equation}
Any displacement $\bm{x}_0$ causes an intensity modulation proportional to
\begin{equation}\label{eq:beta_def}
    d\beta = \frac{8 \pi \rho}{\lambda} \left(\hat{\bm{n}}\cdot\hat{\bm{x}}_0\right) dp_{\textrm{dip}} \; ,
\end{equation}
where $\hat{\bm{x}}_0$ is the unitary vector $\hat{\bm{x}}_0 = \bm{x}_0/\lVert\bm{x}_0\rVert$. The sensitivity to position per unit solid angle, $d\beta$, combines the modulations due to the homodyning field and the distribution of the scattered light. For small absolute displacements $\lVert \bm{x}_0 \rVert \ll \lambda$, the power $dp_{\textrm{det}}$ described by Eq.~\ref{eq:dpdet} is dominated by the term $\rho^2+1$. This constant bias in the signal determines the power spectral density of the fluctuations $d\sigma$ via the shot noise:
\begin{equation}\label{eq:sigma_def}
    d\sigma = \frac{\hbar c}{\lambda}\left(R+1\right)dp_{\textrm{dip}} \; ,
\end{equation}
where $R=\rho^2$ is the reflection coefficient of the mirror for light intensity. Equation~\ref{eq:sigma_def} is valid in experiments in which $\lVert \bm{x}_0\rVert\ll\lambda$, as demonstrated in Ref.~\cite{Cerchiari2020}.
The imprecision in detecting the position $\bm{x}_0$ under a solid angle $d\Omega$ is defined as \cite{Tebbenjohanns2019}:
\begin{equation}\label{eq:differential_imprecision}
    s\left(\hat{\bm{x}}_0\right)=\frac{d\sigma}{d\beta^2}= \frac{\hbar \lambda c \left(R+1\right)}{64 \pi^2 R \left(\hat{\bm{n}} \cdot \hat{\bm{x}}_0\right)^2 dp_{\textrm{dip}}} \; .
\end{equation}
The imprecision captures the interplay between the modulation of the homodyne interference and the radiated power. In an optomechanics experiment, the imprecision corresponds to the noise floor of the detected motional spectrum. Decreasing the imprecision increases the visibility of the thermal mechanical peak of the particle oscillator above the detection noise~\cite{Aspelmeyer2014}. In the context of feedback cooling \cite{Genes2008}, a lower imprecision in the detection of motion is beneficial for reducing the thermal phonon occupation number. The total imprecision over a detector area $D$ is defined as \cite{Tebbenjohanns2019}:

\begin{equation}\label{eq:total_imprecision}
    S_D\left(\hat{\bm{x}}_0\right) =\left(\int_D s\left(\hat{\bm{x}}_0\right)^{-1}\right)^{-1} \; .
\end{equation}

Repeating the calculation from Eq.~\ref{eq:dpdet} to Eq.~\ref{eq:differential_imprecision} for a dipole scatterer measured with the ideal homodyne configuration, we obtain:
\begin{equation}\label{eq:diff_imprecision_ratio}
    \frac{s^{\textrm{\textrm{ideal}}}\left(\hat{\bm{x}}_0\right)}{s\left(\hat{\bm{x}}_0\right)} =
    \frac{4R}{1+R} \;.
\end{equation}
Due to the disparity in the solid angle available for detection, the ratio of total imprecision in the two schemes is only half of the value in Eq.~\ref{eq:diff_imprecision_ratio}:
\begin{equation}\label{eq:Sratio}
    \frac{S^{\textrm{\textrm{ideal}}}_\Omega\left(\hat{\bm{x}}_0\right)}{S_{\Omega/2}\left(\hat{\bm{x}}_0\right)} = \frac{2R}{1+R} \; .
\end{equation}
Since mirrors with reflectivity $>0.99$ are available on the market, we will assume $R\sim1$ in the remainder of this work. This result holds for any orientation of the half-cavity and shows that for very high mirror reflectivity, the half-cavity scheme has the same total imprecision as the ideal configuration in any displacement direction of the dipolar emitter. Thus, we see that the half-cavity setup is a practical implementation of the ideal theoretical configuration and, like the ideal configuration, is limited only by the back action. Note that displacements along different axes are not necessarily detected with the same imprecision. For example, for linearly polarized dipoles, the lowest imprecision $S_{\textrm{min}}=5\hbar c \lambda/(32\pi^2P)$ is achieved for scatterer displacements orthogonal to the polarization direction, which can be compared with the maximal imprecision $S_{\textrm{max}}=5\hbar c \lambda/(16\pi^2P)$ corresponding to displacements along the polarization axis.

\section{Setup with limited NA and a quadrant detector}
In typical levitated optomechanics experiments \cite{Millen_2020}, the scatterer is a SiO$_2$ nanoparticle that oscillates in a harmonic potential generated by optical tweezers or a Paul trap. Each particle is composed of $\sim10^9$~atoms and has a radius of $\sim100$~nm. The trapping frequencies of such particles are $10^2-10^5$~Hz, and the amplitudes of single quanta of oscillation of the particles are $\lVert \bm{x}_0\rVert\sim 10-100$~pm. These excursions are much smaller than the wavelength $\lambda\sim500-1500$~nm of the laser field that illuminates the scatterer, validating the assumption we made in the previous section.

The particle's position is observed via homodyne detection by locating the scatterer between a pair of confocal lenses \cite{Gittes1998,Dawson_19}. The first lens focuses an illuminating Gaussian beam on the particle, and the second lens collimates the scattered light and the incident beam, which then impinge on a detector. In this way, the Gaussian beam is the source for the scattered light and can be used as the reference field for homodyning. This technique is referred to as the ``forward'' detection method. Alternatively, the back-scattered light from the particle can be used to infer the scatterer's position \cite{Felix_asimmetry}. In this case, the returning light is not lost but collimated by the lens that focuses the Gaussian beam onto the nanoparticle. The returning field is interfered with a reference beam for the analysis of the scatterer's position. This technique is called the ``backward'' detection method.

\begin{figure}
\centering
\includegraphics[width=1\columnwidth]{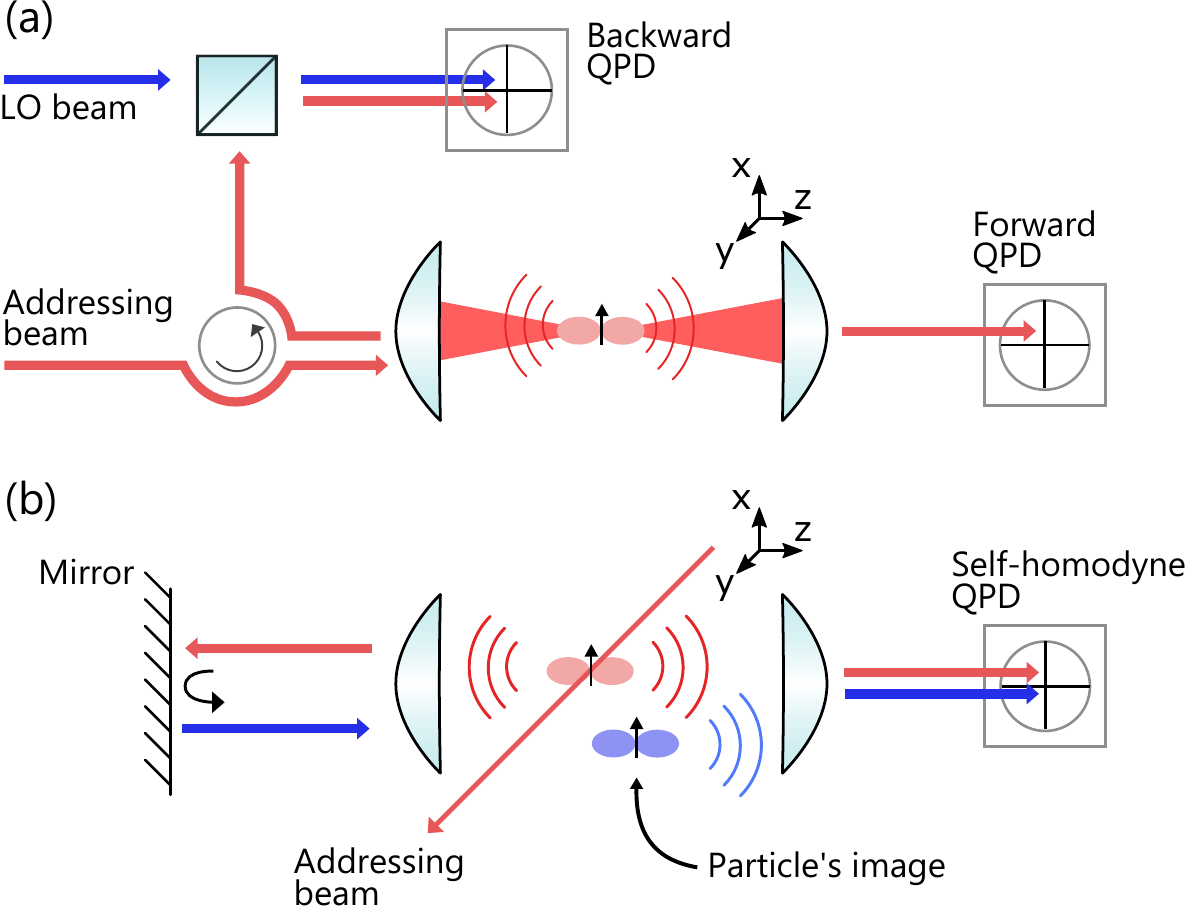}
\caption{Comparison of the forward and backward techniques with the self-homodyne method. (a) Forward and backward techniques. A Gaussian beam propagating along the $z$-axis is focused on the nanoparticle. The scattered light from the particle is detected in the forward and backward directions on quadrant photodetectors (QPD) to measure the scatterer's position. (b) Self-homodyne technique. The half-cavity axis is aligned with the $z$-axis, and the particle is illuminated by a beam propagating along the y-axis. In the main text, we compare the techniques under the assumption that the polarization of the incident light field is aligned with the $x$-axis.}
\label{fig:formicone}
\end{figure}

We present in Fig.~\ref{fig:formicone} a schematic overview of forward, backward and self-homodyne setups to illustrate the configurations that are the basis for the comparison. The forward, backward and self-homodyne detection methods are not mutually exclusive. For example, the forward and backward methods can be used together, as depicted in Fig.~\ref{fig:formicone}(a), with the optical axis oriented along the $y$-axis of Fig.~\ref{fig:formicone}(b). In this configuration the three techniques can be used simultaneously on the same setup. In our comparison of these techniques, we will consider experiments implementing only one technique at a time. This is equivalent to restricting the measurements to having a single detector.

The self-homodyne approach has three advantages compared to the forward and backward setups. First, the mirror generates a secondary dipole image that matches the wavefronts of the primary field, while the Gaussian illuminating beam of the forward technique does not match the wavefronts of the light emitted by the dipolar scatterer. Second, in the self-homodyne method, the coupled system of the mirror and the scatterer emits light in only half of the solid angle. In contrast, both the forward and backward methods may analyze only half of the emitted light scattered by the particle over the entire solid angle. Third, unlike in the forward and backward detection schemes, the optical axis of the half-cavity may have arbitrary orientation with respect to the illumination beam.

In this section, we will show how to make use of the three advantages of self-homodyning to obtain a lower imprecision in the measurement of the scatterer's position than is achievable by means of the forward and backward techniques. In the discussion, we compare the self-homodyne method with the state-of-the-art forward and backward techniques, considering a possible implementation adopting limited-NA optical elements and a quadrant photodetector.

Similarly to the forward and backward methods, the self-homodyne method can be implemented by locating the scatterer between a pair of confocal lenses in order to use a flat mirror and a flat detector, as demonstrated in Refs.~\cite{Slodi2012, Cerchiari2020}. One lens collimates the scattered radiation on a flat mirror for reflection in order to generate the image dipole. The second lens focuses the scattered light and the reference light on the detector. This configuration maps the far-field interference between the primary field and the reflected field onto the detector, a viable solution to reproduce the far-field condition while utilizing a millimeter-size flat detector. In fact, to our knowledge, hemispherical pixel detectors are not yet available. The lenses introduce a cut over the solid angle, which can be modeled by reducing the effective NA of the optical spherical components, as already suggested in Fig.\ref{fig:halfcavity_drawing}.

Three directions describe the setup: the versor $\hat{\bm{z}}$ pointing from the mirror to the detector of the half-cavity system, the displacement of the scatterer $\hat{\bm{x}}_0$ and the polarization $\hat{\epsilon}$ of the illuminating beam. For the description of directions and positions in space, we will make use of angular and Cartesian notation according to the map $(x,y,z) = r(\sin{\theta}\cos{\phi}, \sin{\theta}\sin{\phi}, \cos{\theta})$ as depicted in Fig.~\ref{fig:halfcavity_drawing}(a). We set the axis of the half-cavity along the $z$-axis so that the detector $D$ is located between the polar angles $0<\theta<\theta_D$ at any azimuth $0<\phi<2\pi$. The direction $\hat{\bm{x}}_0$ is indicated by the polar angle $\theta_0$ and the azimuthal angle $\phi_0$, and the polarization $\hat{\bm{\epsilon}}$ by the angles $\theta_\epsilon$ and $\phi_\epsilon$. This configuration can be interpreted in terms of numerical aperture via the expression $\textrm{NA} = \sin{\theta_D}$. To evaluate our hypotheses and prepare the results reported here, we have written a script in Wolfram Mathematica v.~11.3~\cite{Mathematicafile} that calculates the imprecision for arbitrary angles $\theta_0$, $\phi_0$, $\theta_\epsilon$, $\phi_\epsilon$ and $\theta_D$. The solution is cumbersome but useful to evaluate specific experimental implementations where different constraints are present or particular optima of operation are desired. Here, we have decided to focus on the configuration that obtains minimum differential imprecision for a symmetric lens setup, assuming linearly polarized light. In general, a superposition of circular or linear polarizations can be used to illuminate the nanoparticle. However, to minimize the imprecision, linear polarization is preferred because the distribution of scattered light is less homogenous over the solid angle, thus allowing higher intensity to be directed onto a limited-NA detector. In addition, circularly polarized light provides a torque, causing uncontrolled spinning of the levitated nanoparticles~\cite{Kane2010,Reimann2018}. We include the discussion of arbitrary polarization to Appendix C because it is relevant for nonlinear dipolar scatterers such as isolated atoms or molecules.

The configuration with which the minimum differential imprecision $s$ is obtained is found by minimizing Eq.~\ref{eq:differential_imprecision}. We note that the minimization of $s$ with respect to the displacement direction $\hat{\bm{x}}_0$ is independent from the minimization over the light polarization $\hat{\bm{\epsilon}}$. The lowest differential imprecision is found for $\theta_\epsilon =\pi/2$ and  $\phi_\epsilon=0$, corresponding to polarization along the $x$-axis, and for $\theta_0=0$, corresponding to displacement along the $z$-axis. The propagation direction of the illuminating beam is fixed in the $y$z$-$plane by the minimization of $s$. For simplicity, we set the beam orientation to be along the y-axis. In this configuration, the total imprecision as a function of the detector aperture $\theta_D$ is
   
\begin{equation}\label{eq:self_y_NA}
    S_l=\frac{20 \hbar c \lambda}{P \pi^2\left(128-90\cos{\left(\theta_D\right)}-35\cos{\left(3\theta_D\right)} - 3 \cos{\left(5\theta_D\right)}\right)} \; .
\end{equation}

In state-of-the-art experiments, the scattered and reference fields are collimated by a lens on the small detection area of a single detector~\cite{Bushev2006} or a quadrant photodiode~\cite{Tebbenjohanns2019experiment, dania2020}. With the help of these detectors, the particle position is reconstructed by averaging the interference between the scattered and reference fields over large sections of the solid angle which are imaged on the sensitive pixels by the lens. In self-homodyne detection, the intensity common to all quadrants is influenced by the proximity of the scatterer to the detector, and the modulations between pairs of different pixels are caused by displacements of the scatterer in directions orthogonal to the light propagation~\cite{Gittes1998}. The limited angular resolution of the pixels prevents a differential inverse-square weighting of the imprecision to calculate the scatterer's position from the measured intensity variation, as suggested in Eq.~\ref{eq:total_imprecision}. With a discrete pixel detector, the imprecision must instead be calculated as the weighted mean over the available detector area rather than by an integral. This correction has already been presented in Ref.~\cite{Tebbenjohanns2019} to model the forward and backward techniques, and we apply a similar analysis for the self-homodyne method.

\begin{figure}
\centering
\includegraphics[width=1\columnwidth]{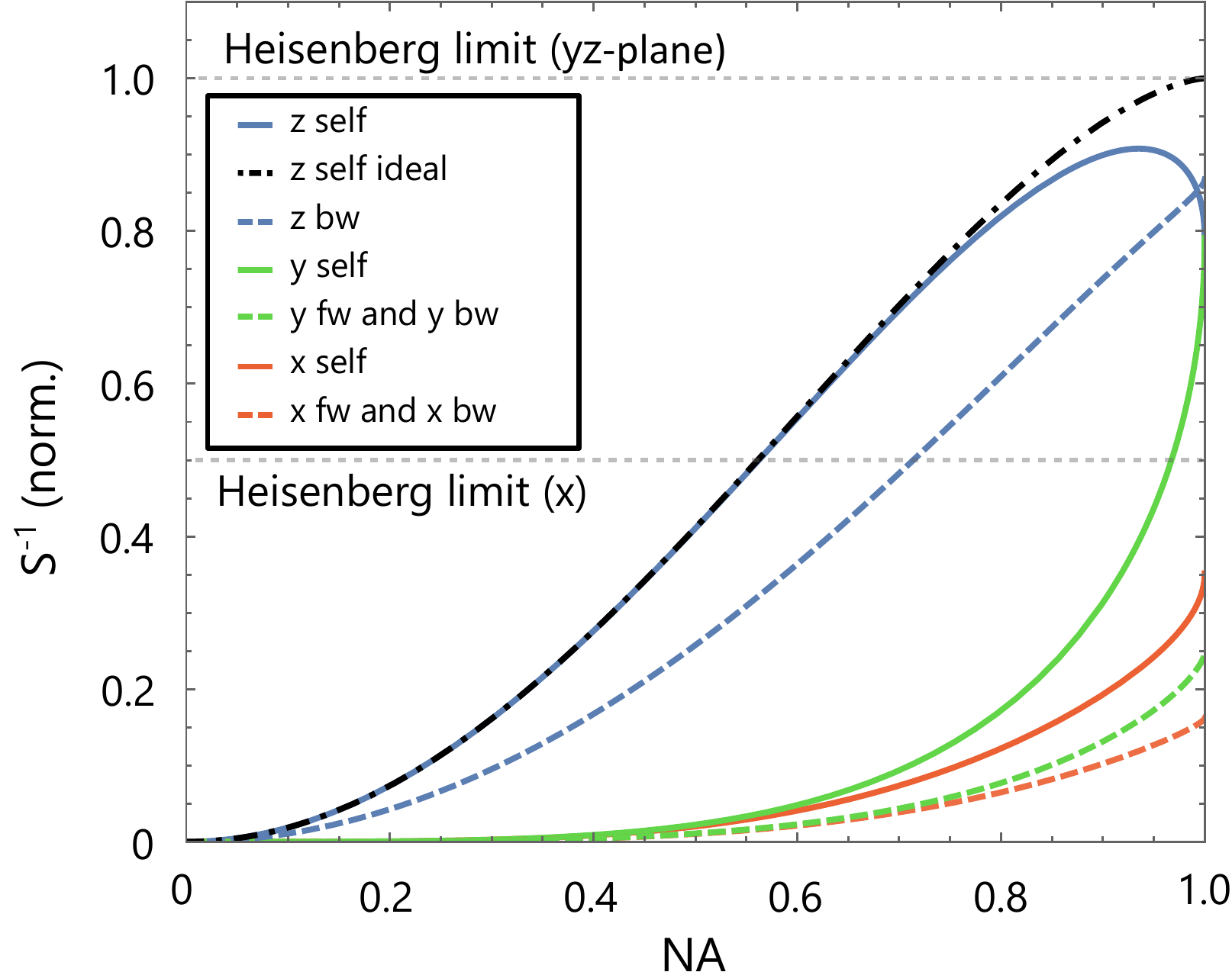}
\caption{Comparison of detection efficiency between the half-cavity setup (\textit{self}) and the forward (\textit{fw}) and backward (\textit{bw}) detection schemes as a function of the NA. The labels $x$, $y$ and $z$ refer to specific motional axes of the scatterer. The forward and backward imprecisions are plotted with dashed colored lines (adapted from \cite{Tebbenjohanns2019}), and the self-homodyne imprecisions with solid colored lines. The lines \textit{fw} and \textit{bw} for the $y$ and $x$ axes overlap. Along the $x$ and $z$ axes, the data obtained with the forward scheme are equal to those obtained with the backward scheme. The efficiency has been normalized for the maximum efficiency at the Heisenberg limit $S^{-1}_{\textrm{min}} = 32\pi^2P/(5 \hbar c \lambda)$. The efficiency for $z$-displacements for the forward method is zero in the selected configuration, and it is not reported in the plot.}
\label{fig:eta_self_vs_gaussian_lin}
\end{figure}

We finally have the tools in hand to compare the forward, backward and self-homodyne methods for experimental application. In this comparison, we align the optical axis of the forward and backward detection along the $z$-axis and the light is linearly polarized along the $x$-axis. For each technique, we calculate the detection efficiency as the inverse of the imprecision in detecting the scatterer displacements along the $x$, $y$, and $z$ axes, and we normalize it with the maximum efficiency obtained at the Heisenberg limit achievable in the system: $S^{-1}_{\textrm{min}} = 32\pi^2P/(5 \hbar c \lambda)$. The efficiencies are calculated as being measured by an ideal quadrant photodetector with unit quantum efficiency. The latter is oriented with its quadrants being divided by the $x$ and $y$ axes.

Figure~\ref{fig:eta_self_vs_gaussian_lin} shows the detection efficiencies of the three techniques as a function of the optical NA of the confocal lenses. The horizontal lines mark the Heisenberg limits for displacements in the $y$z$-$plane and along the $x$-axis. Comparing data for a given axis, the half-cavity shows higher efficiency than the forward or backward detection. We attribute the improvement to both the concentration of the information in only half of the solid angle and the match of polarization and wavefronts between the reference and the scattered light fields. In Fig.~\ref{fig:eta_self_vs_gaussian_lin} we also plot the efficiency corresponding to the ideal pixel detector described by Eq.~\ref{eq:self_y_NA}. A comparison of the ideal detector with the discrete pixel detector, in both cases used in a self-homodyne measurement of displacement along the $z$-axis, reveals that the imprecisions corresponding to the two detectors are equivalent up to $\textrm{NA}\sim0.8$. Thus, we see that the adoption of a quadrant detector is not detrimental for practical implementations aiming to reach the lowest possible imprecision at $\textrm{NA}<0.8$. Further comparisons between the detection efficiencies for displacements along the principal axes acheivable and with a quadrant detector with an ideal detector are presented in Appendix D. Finally, we mention that the freedom of choice in the orientation of the half-cavity optical axis can further be exploited beyond the current configuration, for example, to achieve the imprecision that is shown in Fig.~\ref{fig:eta_self_vs_gaussian_lin} for the half-cavity setup along the $z$-axis along any direction in the $y$z$-$plane.

\section{conclusion}

The self-homodyne detection method implements an ideal setup that is capable of detecting the motion of a dipolar scatterer at the Heisenberg limit. The signal is obtained via self-referencing the scattered light by using a spherical mirror or a combination of a lens and a flat mirror to reflect the light back in the direction of the scatterer. We have analyzed the setup at limited $\textrm{NA}$ for a possible realization that combines two confocal lenses, a flat mirror and a quadrant detector. We have discussed the orientation of the setup that would minimize the imprecision in detecting the scatterer position and that can achieve higher efficiency than the state-of-the-art forward and backward detection methods. Compared to the forward and backward schemes, the mirror generates an ideal reference field and allows for arbitrary orientation of the mirror-detector axis for optimal reconstruction of the scatterer's position by adopting a single one-sided detector. Compared to a deep parabolic mirror optical setup, our method offers a scalable approach for detection that can be adapted in experiments with limited optical access.

In homodyne experiments, the reference may be much stronger than the field being studied, which leads to an amplification of the interference term over the shot noise of a real detector. This is not true in the self-homodyne method, because we expect the primary and the reflected fields to be of similar intensity. Thus, the self-homodyne method requires the adoption of a high quantum efficiency detector. For example, quantum efficiencies larger than 80\% are achievable with avalanche photodiodes in the near-infrared range and even higher efficiencies are possible with superconducting detectors~\cite{Lixing2020}.

In particular, the arbitrary orientation of the half-cavity with respect to the illumination beam constitutes an appealing option for position detection in complex experimental setups. For example, the self-homodyne apparatus could be mounted orthogonally to an optical cavity or to a forward-backward apparatus. Furthermore, the high sensitivity of the half-cavity setup for the  detection of motion at low NAs could be beneficial in levitated optomechanics systems with limited optical access, such as electromagnetic or magneto-gravitational traps.

\hfill \break
\textit{Acknowledgements.} We would like to thank Gabriel Araneda for detailed discussions and suggestions regarding the experimental setup. We thank Lukas Novotny for suggesting us the connection between our method and previous microscopy applications and M. Sondermann for indicating us the studies of dipolar fields in deep parabolic mirrors. G.C. would like to thank Ruggero Caravita for inspiring discussions about isolated systems and about the Wigner-Eckart theorem. 

This work has received funding from the European Union’s Horizon 2020 research and innovation programme under the Marie Skłodowska-Curie grant agreement No 801110 and the Austrian Federal Ministry of Education, Science and Research (BMBWF). It reflects only the author's view, the EU Agency is not responsible for any use that may be made of the information it contains. This work was also supported by the Austrian Science Fund (FWF) Project No. Y951. D.B. acknowledges funding through the ESQ Discovery grant ``Sympathetic detection and cooling of nanoparticles levitated in a Paul trap'' of the Austrian Academy of Sciences. This work has received support by the Institut für Quanteninformation GmbH.

\hfill \break

\renewcommand{\thefigure}{A.\arabic{figure}}
\setcounter{figure}{0}
\renewcommand{\theequation}{A.\arabic{equation}}
\setcounter{equation}{0}
\section{Appendix A: Position of the image}\label{sec:image_position}
In this appendix, we derive that an hemispherical mirror produces a secondary image of a point-like spherical wave emitter on the opposite side of its curvature center. We remind here that the displacement of the emitter from the origin of the half-cavity is small enough to describe the light field propagation by using the Huygens–Fresnel integral. Under this assumption we can model the dipolar scatterer as an emitter of spherical waves. We will show that the mirror generates an image in $\bm{x}_1=-\bm{x}_0$ by demonstrating that the optical path between $\bm{x}_1$ and $\bm{x}_0$ is the same for any part of the outgoing spherical wavefronts which intersect the mirror.

With reference to Fig.~\ref{fig:halfcavity_drawing}, from $\bm{x}_0$, we can follow the wavefront along the generic direction described by the versor $\hat{\bm{v}}_0$ ($\lVert \hat{\bm{v}}_0 \rVert = 1$) until encountering the surface of the sphere of radius $R_s$ at $\bm{x}_r$. There, the wavefront is locally reflected and the light arrives at $\bm{x}_1$ following the direction of the versor $\hat{\bm{v}}_1$. The optical path length between $\bm{x}_0$ and $\bm{x}_r$ is $t_0$ and between $\bm{x}_r$ and $\bm{x}_1$ is $t_1$. For any generic direction $\hat{\bm{v}}_0$ and length $t_1$ the point $\bm{x}_1$ can be found by solving the following system of equations:
\begin{align}
    \label{eq:source_to_mirror}
    \bm{x}_r &=\bm{x}_0 + \hat{\bm{v}}_0 t_0 \\
    \label{eq:radius_mirror}
    \lVert\bm{x}_r\rVert^2 & = R_s \\
    \label{eq:mirror_to_image}
    \bm{x}_1 &=\bm{x}_r + \hat{\bm{v}}_1 t_1 \\
    \label{eq:reflected_versor}
    \hat{\bm{v}_1} &= \hat{\bm{v}}_0 - 2\left(\hat{\bm{v}}_0 \cdot \hat{\bm{x}}_r\right) \hat{\bm{x}}_r \; ;
\end{align}
where $\hat{\bm{x}}_r = \bm{x}_r/\lVert\bm{x}_r\rVert^2$.

The scalar product of Eq.~\ref{eq:source_to_mirror} with $\hat{\bm{v}}_0$ allows us to find the following relation for $t_0$:
\begin{equation}\label{eq:t0}
    t_0 = -\left(\hat{\bm{v}}_0 \cdot \bm{x}_0\right) + R_s \left(\hat{\bm{v}}_0 \cdot \hat{\bm{x}}_r\right) \; .
\end{equation}
Then, we consider the square norm of Eq.~\ref{eq:source_to_mirror}:
\begin{equation}\label{eq:norm_source_mirror}
    R^2_s = \lVert\bm{x}_0\rVert^2 + t^2_0 + 2\left(\hat{\bm{v}}_0 \cdot \hat{\bm{x}}_0\right)t_0 \; .
\end{equation}
By combining Eq.~\ref{eq:t0} and Eq.~\ref{eq:norm_source_mirror}, we obtain
\begin{equation}\label{eq:v0xr}
    \left(\hat{\bm{v}}_0 \cdot \hat{\bm{x}}_r\right) = \sqrt{1-\frac{\lVert\bm{x}_0\rVert^2 - \left(\hat{\bm{v}}_0 \cdot \bm{x}_0\right)^2}{R_s^2}} \;.
\end{equation}
This relation reduces to $\left(\hat{\bm{v}}_0 \cdot \hat{\bm{x}}_r\right) \sim 1$ to first order in $\lVert\bm{x}_0\rVert / R_s$. In this approximation, we find that Eq.~\ref{eq:reflected_versor} reduces to $\hat{\bm{v}}_0\sim\hat{\bm{x}}_r\sim-\hat{\bm{v}}_1$.

With this simplification, we are now ready to evaluate $\bm{x}_0+\bm{x}_1$ by combining Eq.~\ref{eq:source_to_mirror}, Eq.~\ref{eq:mirror_to_image}:
\begin{equation}
    \bm{x}_0+\bm{x}_1 \sim \hat{\bm{x}}_r \left(2 R_s - t_0 - t_1 \right)\;.
\end{equation}
We see that this expression is null if $t_0+t_1 = 2 R_s$. This solution is describing an image formed at $\bm{x}_1=-\bm{x}_0$ because it is valid for any versor $\hat{\bm{v}}_0$. The imaged is formed because $\bm{x}_1$ is the point in which the wavefronts interfere with the same phase delay.
Thus, a point-like light source located near the origin is imaged on the opposite side of the origin with a constant phase delay proportional to twice the radius of the hemisphere.

\renewcommand{\thefigure}{B.\arabic{figure}}
\setcounter{figure}{0}
\renewcommand{\theequation}{B.\arabic{equation}}
\setcounter{equation}{0}
\section{Appendix B: Model description}\label{sec:model_description}

In the self-homodyne method, the light interference of the directly scattered and image fields can be interpreted as a variation of the spontaneous emission rate of the dipolar emitter. The variation of spontaneous emission of the dipolar emitter in half-cavities has already been described in the literature for a perfectly conducting mirror ($\rho=1$) by quantizing the electromagnetic field modes in the presence of a spherical mirror~\cite{Hetet2010}. In this section, we will arrive at the same results by taking into account the interference between the direct field and the reflected field, and we will extend the results of the previous calculation to arbitrary mirror reflectivity.

To calculate the variation of the emission of the dipolar scatterer, we compare the emitted radiation in the presence of the mirror with the emitted radiation in free space. In the presence of the mirror, we divide the solid angle in three regions:
\begin{itemize}
    \item $M$, the angular domain of the mirror,
    \item $\overline{M}$, the symmetric domain to $M$ obtained by reflection through the origin,
    \item $F = 4\pi - M - \overline{M}$, the remaining domain.
\end{itemize}
The ratio of the modified scattering rate $\gamma$ to the free space scattering rate $\gamma_0$ can be calculated as the ratio of emitted powers in the configurations with and without mirror. The ratio references the light exiting at all solid angles from the scatterer-mirror composite system to the light emitted by the scatterer in free space. The expression of the ratio is:
\begin{equation}
    \frac{\gamma}{\gamma_0} = \frac{P\left(M\right)+P\left(\overline{M}\right)+ P\left(F\right)}{P} \; ,
\end{equation}
where $P(D)$ denotes the power radiated in the solid angle $D$ in presence of the mirror and $P$ has already been introduced in Eq.~\ref{eq:power_dipole}. The radiated powers $P(D)$ are
\begin{align}
    P\left(F\right) &= \int_{F} dp_{\textrm{dip}} \; ,\\
    P\left(M\right) &= \int_{M} T dp_{\textrm{dip}}\; ,\\
    P\left(\overline{M}\right) &= \int_{\overline{M}} \left(1+\rho^2- 2\rho\cos{\left(\omega \alpha\left(\hat{\bm{n}}\right)\right)}\right) dp_{\textrm{dip}} \; ,
\end{align}
where $T$ is the transmission coefficient for light intensity. For brevity, we adopt $\alpha\left(\hat{\bm{n}}\right) = \left(\frac{2}{c}\left(\hat{\bm{n}} \cdot \bm{x}_0+R_s\right)\right)$ and $\omega = 2 \pi c/\lambda$. Here, $\rho$ and $T$ can be generic functions of the direction $\hat{\bm{n}}$, but in any direction, $\rho\left(\hat{\bm{n}}\right)^2+T\left(\hat{\bm{n}}\right)=1$, that is, we assume the mirror is lossless. Since for any direction in $M$ there is an opposite in $\overline{M}$, we arrive at:
\begin{align}\label{eq:variation_of_SE}
    \frac{\gamma}{\gamma_0} = 1 -\frac{3}{4\pi}\int_{\overline{M}} \rho\left(-\hat{\bm{n}}\right)\left(1-\lvert\hat{\bm{\epsilon}}\cdot \hat{\bm{n}}\lvert^2\right) \cos{\left(\omega \alpha\left(\hat{\bm{n}}\right)\right) } d\Omega \; .
\end{align}
In the formula, the minus sign in the argument of the $\rho$ function reminds us of the correct reflectivity coefficient while integrating in the $\overline{M}$ region. 
As already pointed out in Ref.~\cite{Hetet2010}, this result indicates that the spontaneous emission of the dipolar emitter can be completely suppressed or doubled in the center of curvature of a hemispherical mirror. Considering the radiated power, we can interpret the underlying mechanism of the self-homodyne scheme in the following way. Assuming perfect reflection, each photon emitted by the scatterer is self-referenced by the mirror, with the result of a position-dependent enhancement or suppression of the spontaneous emission of the dipolar scatterer. Every photon that contributes to the scatterer backaction can only be emitted by the scatterer-mirror system into the open half of the solid angle, and no information is lost on the side of the mirror.

The variation of spontaneous emission determines a modulation of the dipole strength, causing a shift of the energy levels of the dipole. The shift can be calculated from the variation of the scattering rate by using the Kramers-Kronig relations. Applying the Kramers-Kronig formula to Eq.~\ref{eq:variation_of_SE} and approximating the result for a small shift compared to the carrier frequency, we obtain the formula for the energy shift $\Delta\omega$ already found in~\cite{Hetet2010}:
\begin{equation}\label{eq:level_shift}
    \frac{\Delta\omega}{\gamma_0} = -\frac{3}{8\pi}\int_{\overline{M}} \rho\left(-\hat{\bm{n}}\right) \left(1-\lvert\hat{\bm{\epsilon}}\cdot \hat{\bm{n}}\lvert^2\right) \sin{\left(\omega \alpha\left(\hat{\bm{n}}\right)\right) } d\Omega \; .
\end{equation}
Compared to the previous derivation of Eq~\ref{eq:variation_of_SE} and Eq.~\ref{eq:level_shift} reported in Ref.~\cite{Hetet2010}, our approach further identifies the coefficient $\rho$ as the reflection coefficient for the electric field of the mirror.

\renewcommand{\thefigure}{C.\arabic{figure}}
\setcounter{figure}{0}
\renewcommand{\theequation}{C.\arabic{equation}}
\setcounter{equation}{0}
\section{Appendix C: Atomic dipole transition}

The self-homodyne technique can be used to detect the position of an atom. This was demonstrated experimentally for motion along the mirror-detector axis in Refs.~\cite{Cerchiari2020, Bushev2006, Bushev2013, rotter2008monitoring}. Typically, atoms are driven at a frequency close to resonance with a transition between a ground state and an excited state in order to generate light as fluorescence radiation. Under these conditions, the dipolar response of the atom is nonlinear. Unlike a linear dipole, an atom re-emits radiation with both linear and circular polarization regardless of the driving field. Linear polarization corresponds to $\pi$ transitions in which the magnetic quantum number $m$ is the same in the excited state and in the ground state. Circular polarization corresponds to $\sigma$ transitions, i.e., the transition is characterized by a change in quantum number $\Delta m = \pm 1$.

\begin{figure}
\centering
\includegraphics[width=1\columnwidth]{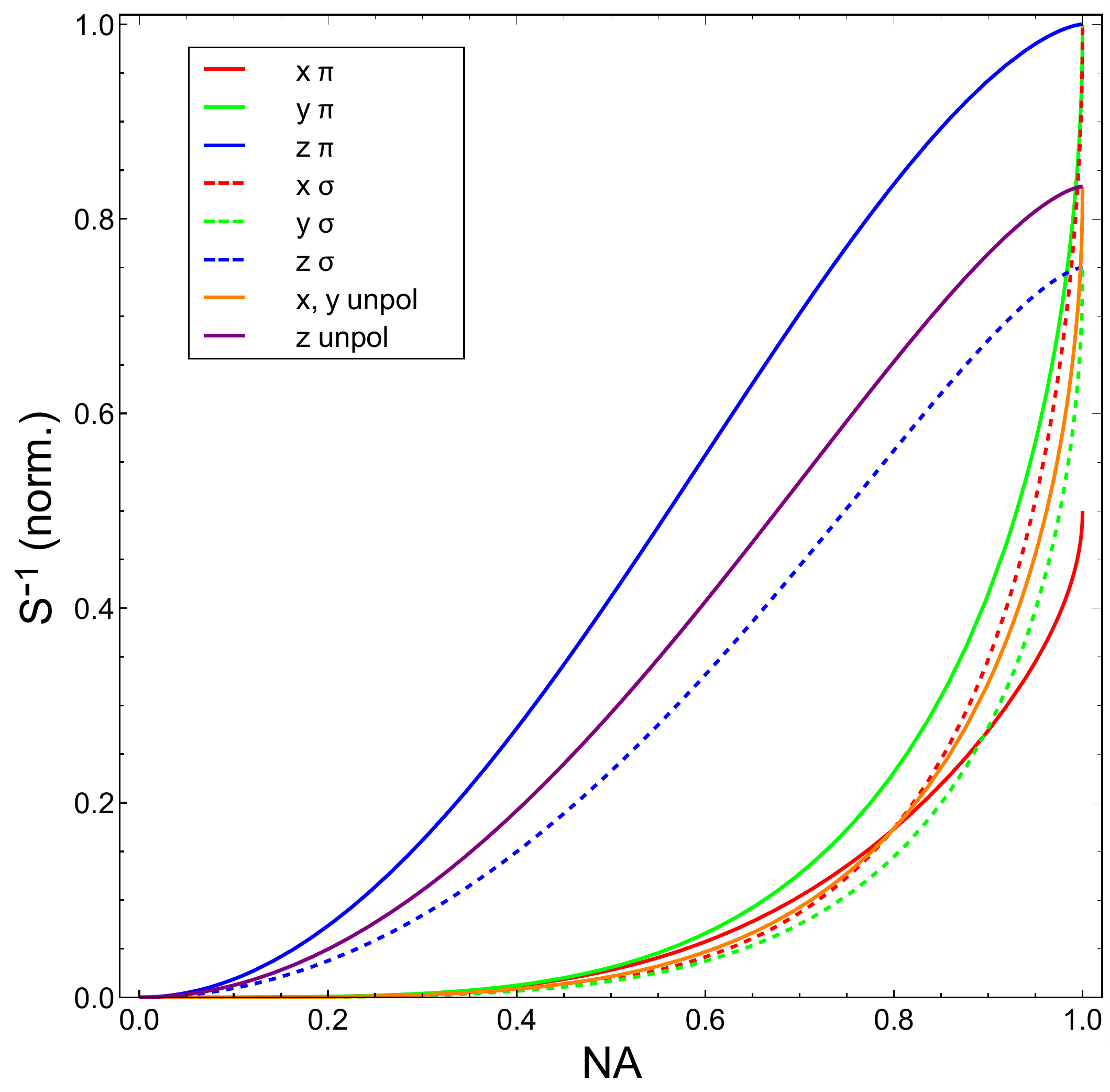}
\caption{Comparison of detection efficiencies for different dipolar scatterer displacements. The labels $\pi$, $\sigma$ and \textit{unpol} (unpolarized) refer to the different angular distributions of emitted radiation as explained in the text. Each curved is also labeled with $x$, $y$ and $z$ according to the axis of displacement of the atom. Unpolarized $x$ and $y$ curves are identical. The values are normalized by $S^{-1}_{\textrm{min}} = 32\pi^2P/(5 \hbar c \lambda)$ as done in the main text.}
\label{fig:imprecision_atom}
\end{figure}

The light emitted during both $\pi$ and $\sigma$ transitions can be used to detect the position of the atom in a self-homodyne setup. Position detection based on $\pi$ transitions is described in the main text. It is equivalent to the detection based on a linear dipolar scatterer driven with linearly polarized light. For compatibility with the result reported in the main text, the quantization axis in an atomic experiment should be aligned along the $x$-axis of Figs.~\ref{fig:halfcavity_drawing} and \ref{fig:formicone} in the main text. In this configuration, in Eq.~\ref{eq:power_dipole}, $\pi$ transitions correspond to light polarized along the axis $\hat{\bm{\epsilon}} = \hat{\bm{x}}$ and $\sigma$ transitions to light polarized along the axes combination described by: $\hat{\bm{\epsilon}} = \left(\hat{\bm{y}} \pm i \hat{\bm{z}}\right)/\sqrt{2}$. Moreover, the unpolarized decay of an atom, defined as the condition in which we do not distinguish between the two transition types but integrate over the possible outcomes, is obtained by the replacement $\left(1-\lvert\hat{\bm{n}}\cdot\hat{\bm{\epsilon}}\rvert^2\right)\rightarrow 2/3$. The last formula can be found by applying the Wigner-Eckart theorem to a generic dipole transition.

We present in Fig.~\ref{fig:imprecision_atom} the efficiencies for the unpolarized, $\pi$-polarized, and $\sigma$-polarized fluorescence of an atom as a function of the system NA. In the plot, we compare the detection efficiencies (inverse imprecisions) for atom displacements along the principal axes of the reference system. In the corresponding calculation, we consider an ideal pixel detector and not a quadradant detector. We see that the detection efficiency depends on the polarization of the scattered light, which is due to the different radiated power distributions for the $\pi$ and $\sigma$ transitions. The displacement of the atom can be detected with highest efficiency along the optical axis of the half-cavity. In this case, the light emitted due to $\pi$ transitions leads to a higher efficiency in the detection of the atom's position than what is achievable by observing $\sigma$-polarized fluorescence. In fact, the distribution of the emitted power is more concentrated in the $yz$-plane in the case of the $\pi$-polarized fluorescence, enabling the reconstruction of the atom position with higher fidelity.

This difference between $\pi$ and $\sigma$ transitions can be tested in experiments by selecting photons based on polarization. The asymmetry can be measured by introducing an external magnetic field along the quantization axis ($x$-axis). If a polarizing beam splitter is used to separate $y$- and $x$-polarized fluorescence, the radiation emitted during $\pi$ and $\sigma$ transitions is separated and can be analyzed independently~\cite{Araneda2019}.
The relevant formulas to reproduce the curves presented in Fig.~\ref{fig:imprecision_atom} can be found by executing the script of Ref.~\cite{Mathematicafile}.

\renewcommand{\thefigure}{D.\arabic{figure}}
\setcounter{figure}{0}
\renewcommand{\theequation}{D.\arabic{equation}}
\setcounter{equation}{0}
\section{Appendix D: Quadrant detector}
In the main text, we compare the imprecision in the reconstruction of the scatterer position using the self-homodyne method with the imprecision obtained by using the forward-backward detection techniques. The comparison takes into account a discrete detector composed of four quadrants. In this section, we describe in more detail the calculation with which we obtain the imprecision of the self-homodyne technique as measured with a quadrant photodetector (QPD), and how it deviates from the idealized model of a differential pixel detector. In the discussion, we assume the same configuration shown in Figs.~\ref{fig:halfcavity_drawing} and \ref{fig:formicone}, in which the QPD occupies the solid-angle regions $Q$ contained in the polar-angle interval $0 \leq \theta \leq \theta_D$. The quadrants regions $Q_n$ split the $Q$ region in the azimuthal-angle intervals $(n-1)\pi/2\leq\phi < n\pi/2$, with $n=\{1,2,3,4\}$. The total power impinging on the $n$-th quadrant is
\begin{equation}
P^{Q_n}=\int_{Q_n}\big(1+\rho^2+\rho\frac{8\pi}{\lambda}(\hat{\bm{n}}\cdot \bm{x}_0)\big)dp_{\textrm{dip}}\; .
\end{equation}
We reconstruct the particle's displacement $\bm{x}_0=(x_0,y_0,z_0)$ by taking advantage of the symmetries of the integrand. The combinations of the detected powers over the quadrants that we use to reconstruct the displacements are~\cite{Tebbenjohanns2019,Gittes1998}:
\begin{equation}
    \begin{split}
        x_0 &= \Big((P^{Q_1}+P^{Q_4})-(P^{Q_3}+P^{Q_3})\Big)\frac{1}{4B(\frac{\pi}{2},0)},\\
y_0 &= \Big((P^{Q_1}+P^{Q_2})-(P^{Q_3}+P^{Q_4})\Big)\frac{1}{4B(\frac{\pi}{2},\frac{\pi}{2})}, \text{~and}\\
z_0 &= \Big(\sum_nP^{Q_n}-\int_{Q}(1+\rho^2)dp_{\textrm{dip}}\Big)\frac{1}{4B(0,\text{any})}.
    \end{split}
\end{equation}
Here, $B(\theta_0,\phi_0) = \int_{Q_n} d\beta$ is the integral of the sensitivity to displacements along $\hat{\bm{x}_0}$ defined in Eq.~\ref{eq:beta_def} and is equal for all quadrants thanks to the detector symmetry. The imprecision measured with the QPD is calculated as 
\begin{equation}
S_{QPD}(\hat{\bm{x}}_0)=\frac{1}{16 \, B(\theta_0,\phi_0)^2} \int_{Q}\sigma d\Omega,
\end{equation}
where $\sigma$ is the spectral density of the power fluctuations defined in Eq.~\ref{eq:sigma_def}. 
\begin{figure}
\centering
\includegraphics[width=1\columnwidth]{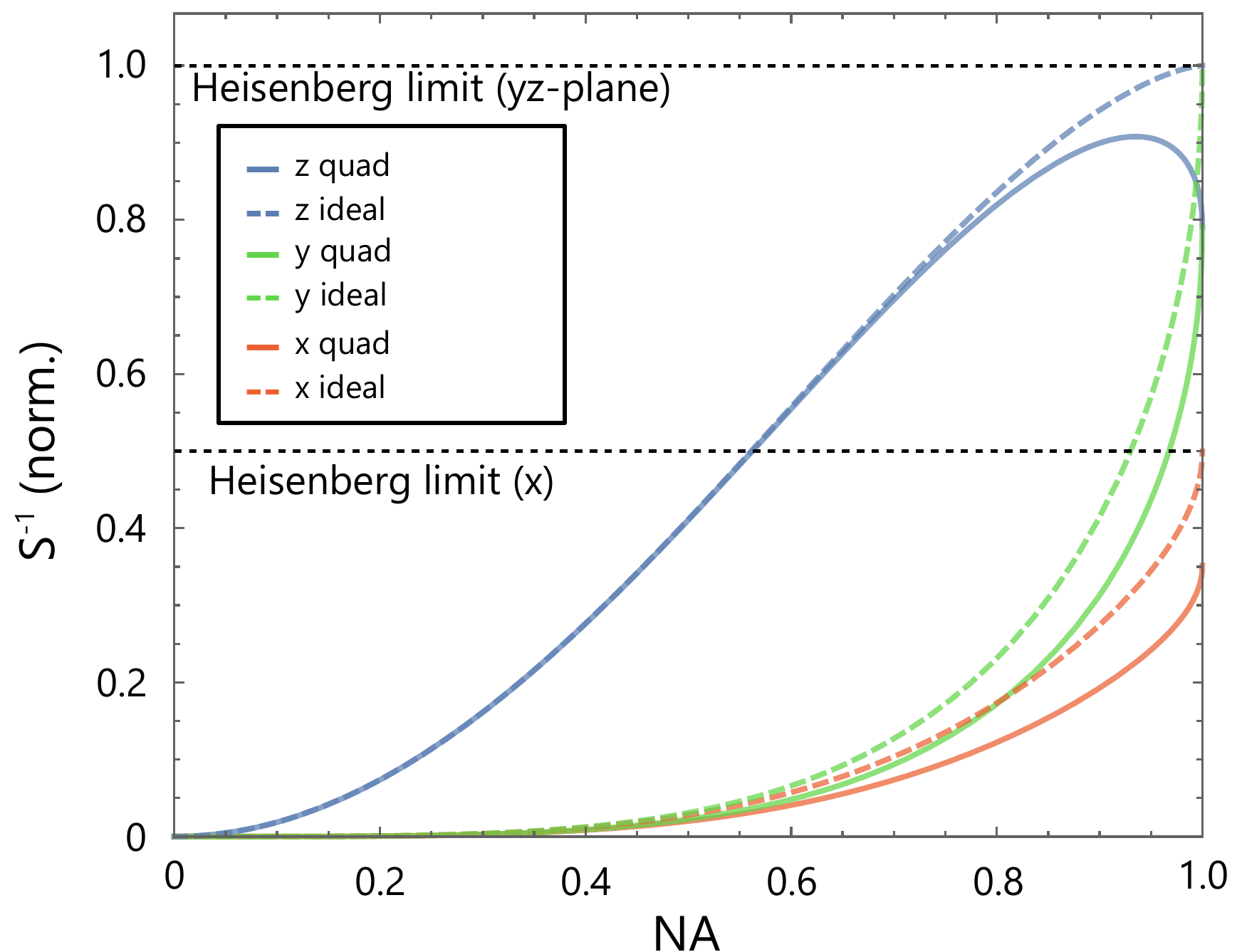}
\caption{Comparison of the detection efficiencies of self-homodyne detection along three axes with an ideal differential pixel detector, in dashed lines, and using a quadrant photo detector, in solid lines, as a function of the detector numerical aperture ($\text{NA}$). Values are normalized with respect to the efficiency at the Heisenberg limit for displacements on the $\phi_0=\pi/2$ plane: $S^{-1}_{\textrm{min}} = 32\pi^2P/(5 \hbar c \lambda)$. }
\label{fig:ideal_vs_real}
\end{figure}

Figure \ref{fig:ideal_vs_real} shows a comparison of $S^{-1}_{QPD}$, the detection efficiency of the self-homodyne method obtained with the QPD (solid lines), and $S^{-1}_D$, the detection efficiency obtained with the ideal pixel detector (dashed lines), as a function of the detector numerical aperture $\text{NA}=\sin(\theta_D)$. Values are normalized by $S^{-1}_{\textrm{min}} = 32\pi^2P/(5 \hbar c \lambda)$ as done in the main text. The expression for the ideal efficiency $S^{-1}_D$ is obtained by integrating Eq.~\ref{eq:total_imprecision} of the main text. We find that detection of the particle's motion along the $z$-axis with the QPD is equivalent to the ideal pixel detector up to $\text{NA}\sim0.8$, suggesting that differential pixel weighting only becomes relevant in the limit $\theta_D \rightarrow \pi/2$. Supporting this hypothesis, the imprecision in detecting motion along the $x$ and $y$ axes deviates from the ideal case already at lower $\textrm{NA}$s. The early NA deviation of the two curves can be understood based on the angular distribution of the interference pattern. For $x$ and $y$ displacements, the intensity is mostly modulated at angles $\theta \sim\pi/2$ and modulated less at angles $\theta \sim 0$ (Eq.~\ref{eq:interference_halfcavity}), i.e., at polar angles for which the effect of differential inverse weighting is more pronounced.

\bibliography{bibliography}

\end{document}